\documentclass{elsart} 

\usepackage[left,pagewise]{lineno}
\usepackage[dvips]{graphicx}
\usepackage[latin1]{inputenc}
\usepackage{amssymb,array,url,amsmath}
\usepackage{picins}
\usepackage{algorithmic}
\usepackage{algorithm}

\journal{Neurocomputing}

\begin{document}
\sloppy
\begin{frontmatter}
  \title{Batch kernel SOM and related Laplacian methods for social network analysis\thanksref{merci}}
  
  \author[rb]{Romain Boulet},
  \author[rb]{Bertrand Jouve},
   \author[fr]{Fabrice Rossi}, and
    \author[nv]{Nathalie Villa}
    \corauth[cor]{Corresponding author.}
  \ead{nathalie.villa@math.univ-toulouse.fr}
  \address[rb]{Institut de Mathématiques, Université de Toulouse et CNRS (UMR 5219), 5 allées Antonio Machado, 31058 Toulouse cedex 9, France}
  \address[fr]{Projet AxIS, INRIA Rocquencourt, Domaine de Voluceau, Rocquencourt, B.P. 105, 78153 Le Chesnay cedex - France}
	\address[nv]{Institut de Mathématiques, Université de Toulouse et CNRS (UMR 5219), 118 route de Narbonne, 31062 Toulouse cedex 9, France}
  
  \thanks[merci]{This work was partially supported by ANR Project ``Graph-Comp''.}
  
  \begin{abstract}
    Large graphs are natural mathematical models for describing the structure
    of the data in a wide variety of fields, such as web mining, social
    networks, information retrieval, biological networks, etc. For all these
    applications, automatic tools are required to get a synthetic view of the
    graph and to reach a good understanding of the underlying problem. In
    particular, discovering groups of tightly connected vertices and
    understanding the relations between those groups is very important in
    practice. This paper shows how a kernel version of the batch Self
    Organizing Map can be used to achieve these goals via kernels derived from
    the Laplacian matrix of the graph, especially when it is used in
    conjunction with more classical methods based on the spectral analysis of
    the graph. The proposed method is used to explore the structure of a
    medieval social network modeled through a weighted graph that has been
    directly built from a large corpus of agrarian contracts.
  \end{abstract}
\end{frontmatter}



\section{Introduction}
\emph{Complex networks} are large graphs with a non trivial organization. They
arise naturally in numerous context \cite{bornholdt_schuster_HGN2003}, such as,
to name a few, the World Wide Web (which gives a perfect example of how large
and complex such a network may grow), metabolic pathways, citation networks
between scientific articles or more general social networks that model
interaction between individuals and/or organizations, etc.

Complex networks share common properties that have allowed the emergence
of mathematical descriptions such as \emph{small world graphs} or \emph{power law
  graphs}. The structure of these graphs often gives some keys to understand
the complex network underlined. To study such a structure, one often begins
with a metrology process applied to the graph that describes the degree
distribution, the number of components, the density, etc. The second step consists in the
search of subgraphs that have particular adjacency, in particular highly
connected parts of the graph which are at the same time lightly connected
between them. Such parts are called \emph{communities}\footnote{It should be
  noted that this use of the term ``community'' while quite standard in
  computer science is disputed in other disciplines, e.g., sociology, where
  a group of individuals highly connected in some sense is not always
  considered as forming a community.}
\cite{pons_latapy_JGAA2006,donetti_munoz_JSMTE2004}. More recently, several
directions have been explored to go further in modelling large real networks,
taking into account their dynamics \cite{strogatz_N2001}, the attributes of
the data \cite{neville_etal_PTMLAW2003}, a more formal definition of the
communities
\cite{radicchi_etal_PNASUSA2004,newman_girvan_PRE2004,newman_etal_SDN2006}, or
the relations between communities \cite{palla_etal_N2005}. However, it should be noted
that dealing with very large graphs (millions of vertices) is still an
open question (see \cite{clauset_etal_PR2004} for an example of an efficient
algorithm to explore that kind of data sets).

Several ways have been explored to cluster the vertices of the graph into
communities \cite{schaeffer_CSR2007} and some of them have in common the
use of the \emph{Laplacian} matrix. Indeed, there are important relationships
between the spectrum of the Laplacian and the graph invariants that
characterize its structure (see,
e.g. \cite{mohar_LSG1991,mohar_poljak_ECO1993}). These properties can be used
for building, from the eigen-decomposition of the Laplacian, a similarity
measure or a metric space such that the induced dissimilarities between
vertices of the graph are related to its community structure (see
\cite{donetti_munoz_JSMTE2004}, among others). The Laplacian matrix also appears
when the vertices of the graph are clustered by the optimization of a graph
cut quality measure: optimizing such a measure is generally a NP-complete
problem but using the properties of spectrum of the Laplacian provides a relaxation based heuristic solution with a reasonable complexity \cite{alpert_kahng_ICDA1993}.

In the present paper, the properties of the Laplacian are also used to identify
and map communities, both in a rather classical way and with a recently
proposed batch version of the kernel Self Organizing Map (SOM). The
combination of those tools gives complementary views of a social network. The
spectral based approach extracts a specific type of communities for which
interpretation biases are limited, but which cover only part of the graph
(e.g., one third of the vertices in the studied graph). The SOM solution gives
a global map of the vertices clustered into more informal communities, for
which a link analysis must be done with care. Combining the analysis of these classifications helps in
getting some global results while limiting the risk of false interpretation. 

In both cases, the communities' organization is associated to a two dimensional representation that eases interpretation of their relations. It should be noted
that these representations are not intended to compete with those coming from
the large field of graph drawing
\cite{dibattista_etal_GD1999,herman_etal_IEEETVCG}: the goal in this paper is
not to draw the whole graph but to extract a community structure and to
provide a sketch of the organization of these small homogeneous social groups.

The rest of this paper is organized as follows. Section \ref{algebric} defines
\emph{perfect communities} and uses spectral analysis of the Laplacian to
identify them.  An alternative and complementary approach is described in
Section \ref{diff_kernel}, where a \emph{kernel} is derived from the Laplacian via a form of regularization. This kernel is used in Section \ref{som} to
implement a batch kernel SOM which builds less perfect communities and maps
them on a two dimensional structure that respects their relationships. Section
\ref{sectionMedievalNetwork} is dedicated to an application of the proposed methods to a social network that models interactions between peasants in the French medieval society. The historical sources (agrarian contracts) are first presented together with the corresponding social network model. Methods proposed in Sections \ref{algebric} and \ref{som} are then applied to this graph. Results are compared and confronted to prior historical knowledge.

\section{Clustering through the search of perfect communities\label{algebric}}

Understanding the structure of a large network is a major challenge. Fortunately, many real world graphs have a non uniform link density: some groups of vertices are densely connected between them but sparsely connected to outside vertices. Identifying those \emph{communities} is very useful in practice \cite{newman_girvan_PRE2004} as they can provide a sort of summary which can in turn be analyzed more easily than the original graph, especially when human expertise is requested. However, there is no consensus on a formal definition of a community (see e.g., \cite{pons_latapy_JGAA2006,schaeffer_CSR2007}). In the context of visualization, a particular (and somehow restrictive) type of communities, the so called \emph{perfect communities}, leads to interesting results. In addition, this precise form of communities, easy to define and understand, may be viewed as the elementary block of the communities, in their general meaning.

A perfect community of a non-weighted graph is a complete subgraph (in such a subgraph all vertices are pairwise linked by an edge), with at least $2$ vertices, and such that all its vertices have exactly the same neighbors outside the community. The perfect communities of a weighted graph are obtained as the perfect communities of its induced non-weighted graph (i.e., of the graph having same vertices and edges but no weights on the edges). In \cite{vandenheuvel_pejic_AOR2001}, Van den Heuvel and Pejic proposed that particular form of community for non-weighted graphs in the case of frequency assignment problems.  They give a different definition for weighted graphs; this last one was not followed because it appears as too restrictive for more general graphs such as those coming from social networks. 

A nice advantage of perfect communities over looser ones is that they have simple non ambiguous visual representations. Indeed perfect communities can be represented by simple glyphs (circles for instance) together with their connections to other perfect communities without loosing information: the nodes in a perfect community are fully connected (hence each simple glyph symbolizes a complete subgraph) and share the same connections with the outside of the community (hence the unique representation of these connections by a simple link between two glyphs).

However, perfect communities don't provide a complete summary of a graph. One of their main weaknesses is that, on real applications, the set of perfect communities can contain only a part (and sometimes a little part) of the whole graph. Moreover, some of the vertices that don't belong to a perfect community can play a central role in the structure of the social network. Two parameters are usually used in social network analysis to characterize these important vertices: high degree and high betweenness measure (the definition is given below).

The vertices with the highest degrees are likely to have a main role in the graph as they are linked to a large number of other vertices. These vertices may appear in a \emph{rich-club} \cite{zhou_mondragon_IEEECL2004} if it exists.
The rich-club occurs when the vertices with highest degree form a dense subgraph with a small diameter.
The diameter of a graph is the longest of the shortest path between any two given vertices of a graph and the density of a graph is the ratio beetwen the number of its edges and the number of the total possible edges. The construction of a rich-club starts from the highest degree vertices which are totally connected and follows by adding the next vertices in the decreasing order of their degrees. The process stops when the diameter reaches the fixed limit or when the density sharply decreases. In practice, the chosen limit for the diameter of the rich club is very small: for a graph having several hundred of vertices, as the one studied in Section~\ref{sectionMedievalNetwork}, a diameter of 2 gives satisfactory results. As the rich club is a subgraph with a small diameter and a high density and as it shares many connections with the other vertices of the graph, it can be seen as a set of people having a main social role by knowing almost everybody in the community.

All the vertices of the graph don't belong to a perfect community or to the rich club and some of them can still be important to obtain a good summary of the graph. Another interesting feature to localize relevant vertices is to look at the \emph{betweenness measure} of the vertices. The betweenness measure of a vertex is the frequency of the shortest paths of any two vertices of the graph in which this vertex occurs. These vertices also have a main role as they are essential to connect the whole graph. In social networks, they can be seen as mediating persons that link together subgroups that would be otherwise unrelated. 

The number of high betweenness vertices is chosen according to the following heuristic. Vertices are sorted in decreasing order of betweenness and the number of connected components of each subgraph $\mathcal{S}_k$ induced by the perfect communities, the rich-club and the first $k$ vertices with highest betweenness measure is computed. In general, the decrease of this number with $k$ is non uniform: sharp drops are separated by constant (flat) regions (see Figure \ref{riche_et_central} for an example). As important vertices are those that significantly reduce the number of connected components, it seems logical to consider a value of $k$ that lies just after a significant drop. The actual selection of $k$ remains however a matter of compromise as there are generally several significant decreases in the number of components: adding too much nodes will clutter the visualization while leaving out too many will miss some important individuals. Section \ref{subsec:medieval:perfect} provides an example of such compromise. The final set of selected vertices are called \emph{central vertices}.

Adding the rich-club and the central vertices to perfect communities enhances the coverage of the original graph while maintaining an easy visual representation. The first step consists in using adding glyphs for central vertices. As already explained above, links from a perfect community to any vertex is unambiguous and therefore the edges between perfect communities and central vertices (as well as between those vertices) can be added without difficulty. The only compromise concerns the rich-club. It is also represented by a specific glyph which does not show therefore its substructure. Another simplification is used for links: an edge between the rich-club and any other element (a central vertex or a community) summarizes a possibly complex link structure. 

Perfect communities are not only easy to visualize; their computation is also
straightforward, as described below. Let us first introduce some notations. $\mathcal{G}$ denotes a connected graph with vertices $V=\{x_1,\ldots,x_n\}$ and a set of undirected edges $E$, with positive weights, $w_{i,j}=w_{j,i}$ ($w_{i,j}=0$ is
equivalent to $\{x_i,x_j\}\notin E$). The degree of a vertex $x_i$ is denoted $d_i=\sum_{j=1}^n w_{i,j}$. 

The structure of $\mathcal{G}$ can be summarized through a symmetric $n\times n$
matrix called the \emph{Laplacian} of ${\mathcal{G}}$. This matrix has been
intensively studied the past years because many important structural and
topological properties can be deduced from it.  The Laplacian of
${\mathcal{G}}$ is defined as the positive and semi-definite matrix
$L=(L_{i,j})_{i,j=1,\ldots,n}$ such that
	\[
	 L_{i,j} = \left\{ \begin{array}{ll}
		-w_{i,j} & \textrm{ if }i\neq j,\\
		d_i & \textrm{ if }i=j.
	\end{array}\right. 
	\]
We will also consider the Laplacian, denoted by $\widetilde{L}$, of the non-weighted graph induced by $\mathcal{G}$, $\widetilde{\mathcal{G}}$.

Spectral properties of the Laplacian can be used to cluster the vertices of a
graph. First of all, it is well known that the eigenvalue 0 is related to the
minimum number of connected subgraphs in ${\mathcal{G}}$
\cite{mohar_LSG1991}. In the same way, a spectral analysis of the Laplacian allows one to find 
 perfect communities, using the following property (a set of vertices is called non-stable if it contains at least two adjacent vertices): 
\begin{thm}[\cite{vandenheuvel_pejic_AOR2001}]\label{vandenheuvel}
	A non-stable set $\mathcal{S}$ of vertices is a perfect community if and only if there is a non-zero eigenvalue, $\lambda$, of $\widetilde{L}$ whose multiplicity is at least $k-1$ and such that the $k-1$ associated eigenvectors vanish for the same $n-k$ coordinates.

	Then, the cardinal of $\mathcal{S}$ is $k$, the coordinates for which the $k-1$ eigenvectors are not 0 represent the vertices belonging to $\mathcal{S}$ and $\lambda=d+1$ where $d$ is the degree of a vertex of $\mathcal{S}$.
\end{thm}

As a consequence, looking at null coordinates of the eigenvectors of $\widetilde{L}$ is a simple and efficient way to extract perfect communities. Moreover, we also have the following property, that will help to understand the link between this approach and the well-known spectral clustering method:
\begin{cor}
	\label{eigen_communaute}
	If a set of vertices, $\mathcal{S}$, is a perfect community then the $n-k+1$ eigenvectors that do not define $\mathcal{S}$ have constant coordinates for the indices of the vertices of $\mathcal{S}$.
\end{cor}

{\bf Proof} \qquad Without loss of generality, ${\mathcal{S}}$ is renumbered as
${\mathcal{S}}=\{1,\ldots,k\}$. An eigenvector $u$ defining $\mathcal{S}$ can be
written as $u=(u_1,...,u_k,0,0,...,0)$. Let $z=(z_1,...,z_k,z_{k+1},...,z_n)$
be an eigenvector that does not define $\mathcal{S}$ and note
$\tilde{u}=(u_1,...,u_k)$, $\tilde{z}=(z_1,...,z_k)$. As $\widetilde{L}$ is
symmetric, $z$ is orthogonal to the $k-1$ eigenvectors that define
$\mathcal{S}$ so $\tilde{z}$ is orthogonal to the $k-1$ vectors $\tilde{u}$
for $u$ defining $\mathcal{S}$. But $u$ is an eigenvector of $\widetilde{L}$, so it
is orthogonal to the vector $\mathbf{1}_n=(1,1,\ldots,1)$ (related to the
eigenvalue 0) and so, $\tilde{u}$ is orthogonal to the vector
$\mathbf{1}_k$ in $\mathbb{R}^k$. The orthogonal complement, in
$\mathbb{R}^k$, of the vector space spanned by the $k-1$ vectors $\tilde{u}$
has dimension one and is spanned by $\mathbf{1}_k$; it follows that
$\tilde{z}$ is co-linear to $\mathbf{1}_k$ which concludes the proof.

\section{Similarity measures built from the Laplacian\label{diff_kernel}}
Some weaknesses of a representation by perfect communities are the absence of
a lot of vertices (for instance, only 35 \% of the whole graph belongs
to a perfect community in the social network studied in Section \ref{sectionMedievalNetwork}) and the presence of a lot of very small communities.
Moreover, some relevant groupings of
perfect communities might be missed and a bias of the interpretation can occur
from these lacks. In this sense, the definition of perfect communities gives a too restrictive clustering of the vertices. It is therefore reasonable to complement it with the help of another clustering algorithm chosen in the numerous methods proposed for this task \cite{schaeffer_CSR2007}. A broad class of those methods consists in building a (dis)similarity measure between vertices that capture the notion of community and then on applying an adapted clustering algorithm to the dissimilarity matrix. 

To pursue this goal, this section introduces existing similarity measures
based on the Laplacian. Section \ref{cutoff} explains how to build a
similarity measure that is able to separate communities from each others and
Section \ref{heat_kernel} follows a similar idea to define a \emph{kernel}
that maps the vertices in a high dimensional space. The purpose of the
those sections is to emphasize the links and also the differences between
the eigenvalue approach described in the previous section, the usual
``spectral clustering'' approach and the well-known \emph{diffusion kernel}
which can be considered as a smooth spectral clustering.

\subsection{From almost perfect communities to graph cuts\label{cutoff}}
One way to obtain an optimal clustering of the vertices of a graph is to
minimize the following \emph{graph cut} quality measure
\[
\text{cut}({\mathcal{S}}_1,\ldots,{\mathcal{S}}_p)=\sum_{i=1}^p W({\mathcal{S}}_i,{\mathcal{G}}\setminus{\mathcal{S}}_i),
\]
where ${\mathcal{S}}_1,\ldots,{\mathcal{S}}_p$ is a partition of $V$ (for a
chosen $p$ in $\mathbb{N}^*$), and where $W(S,S')=\sum_{i \in S, j\in S'}
w_{i,j}$ for two given sets of vertices $S$ and $S'$ included in $V$. This
optimization problem is NP complete for $p >2$ but it can be \emph{relaxed}
into a simpler problem (see, e.g., \cite{vonluxburg_TR2007}):
\begin{equation}
\label{min_cut_pb}
\min_{H\in \mathbb{R}^{n\times p}} \text{Tr}(H^TLH)\text{ subject to }H^TH=\mathbb{I}_p.
\end{equation}
The key point in the relaxation approach is to extend the search space from a
discrete set in which the coefficients of $H$ define a partition of
$\{1,\ldots,n\}$, to $\mathbb{R}^{n\times p}$. 

For a connected graph, the solution of the relaxed problem is the matrix $H$
which contains the $p$ eigenvectors associated to the $p$ smallest positive
eigenvalues of $L$ as columns. Of course, the real-valued solution provided by
the matrix $H$ has to be converted into a discrete partition of $p$
clusters. A usual way to do do is to consider the solution matrix $H$ as a way
to map vertices of the graph in $\mathbb{R}^p$ as follows:
\begin{equation}
\label{F_L}
	F_L:x_i \in V \rightarrow (h^{(1)}_i,\ldots,h^{(p)}_i) \in \mathbb{R}^p,
\end{equation}
where $(h^{(j)})_j$ is an orthonormal set of eigenvectors associated to the
$p$ smallest positive eigenvalues $(\lambda_j)_{j=1,\ldots,p}$ ($h^{(j)}_i$
denotes the $i^\text{th}$ coordinate of the $j^\textrm{th}$ smallest
positive eigenvalue). Then a standard clustering algorithm in $\mathbb{R}^p$ (e.g., the $k$-means algorithm) is applied to the mapped nodes (this can been seen as a clustering of the rows of $H$), leading to one variant of \emph{spectral clustering}. 

This method is strongly related to perfect communities calculation. Corollary~\ref{eigen_communaute} shows that vertices that belong to the same perfect
communities have the same coordinates for many eigenvectors. As a
consequence, any clustering algorithm applied on nodes mapped via $F_L$ will
tend to gather vertices from a perfect communitiy in the same cluster. In
this sense, spectral clustering can be seen as a relaxed version of the search
of perfect communities. 

It should be noted that the spectral clustering method summarized above gives
equal weights to the first $p$ eigenvectors of the Laplacian, whereas the
smaller the eigenvalue is, the more important the corresponding eigenvector
is. Moreover, only the first $p$ eigenvalues are used and, hence, this approach doesn't use the entire information provided by the Laplacian. To avoid these problems, a regularized version of the Laplacian can be used, as shown in the following section.

\subsection{Diffusion kernel\label{heat_kernel}}
In \cite{smola_kondor_COLT2003}, the authors investigate a family of kernels
on graphs based on the notion of regularization operators: a regularization
function is applied to the Laplacian and gives a family of matrices that are
also kernels on $V\times V$. In the present paper, we focus on the
\emph{diffusion kernel}:
\begin{defn}
	The diffusion matrix of the graph ${\mathcal{G}}$ for the parameter $\beta>0$ is $D^\beta=e^{-\beta L}$.
	
	The diffusion kernel of the graph $\mathcal{G}$ is the function
		\[
		K^\beta:(x_i,x_j)\in V\times V\rightarrow D^\beta_{i,j} \in \mathbb{R}
		\]
\end{defn}
The diffusion matrix is easy to compute for graphs having less than a few hundred of vertices by the way of an eigen-decomposition of the Laplacian: if $(h^{(i)})_{i=0,\ldots,n-1}$ are orthonormal eigenvectors associated to the eigenvalues $0=\lambda_0 < \lambda_1 \leq \lambda_2 \leq \ldots \leq \lambda_{n-1}$ of $L$, then 
\begin{equation}
\label{diffusion_kernel}
	D^\beta=\sum_{k=0}^{n-1} e^{-\beta \lambda_k} h^{(k)} {h^{(k)}}^T.
\end{equation}

This diffusion kernel has been intensively studied through the past years. In particular, \cite{kondor_lafferty_ICML2002} shows that this kernel is the continuous limit of a diffusion process on the graph: $K^\beta(x_i,x_j)$ can be viewed as the value of the energy obtained in vertex $x_j$ after a time tending to infinity if energy has been injected in vertex $x_j$ at time 0 and if diffusion is continuously done among the edges of the graph. In this case, $\beta$ is related to the intensity of the diffusion (see also \cite{chung_SGT1997} for a complete description of the properties of this operator).

It is easy to prove that the kernel $K^\beta$ is symmetric and definite
positive. Then, from Aronszajn's Theorem
\cite{aronszajn_TAMS1950,berlinet_thomasagnan_RKHSPS2004}, there is a
Reproducing Kernel Hilbert Space (RKHS), $({\mathcal{H}}_\beta,\langle.,.\rangle_\beta)$, called the feature space, and a mapping
function, $\phi_\beta:V\rightarrow {\mathcal{H}}_\beta$ such that: 
\[
	\textrm{for all }i,j,\ \langle \phi_\beta(x_i),\phi_\beta(x_j)\rangle_\beta = K^\beta(x_i,x_j).
\]
As in the previous section, this mapping provides a way to apply standard
clustering algorithms to the vertices of a graph, simply by working on their
mapped values. In addition, a \emph{kernel trick} can be used in many cases to
avoid calculating explicitly the mapping (see Section \ref{BatchKernelSOM} for
details). 

Equation \eqref{diffusion_kernel} shows that the mapping induced by
the kernel is equivalent to the following one:
\begin{equation}
	\label{F_K}
	F_K^\beta : x_i \in V\rightarrow (h^{(1)}_i,\ldots,h^{(n)}_i) \in (\mathbb{R}^n,\langle.,.\rangle_n^\beta),
\end{equation}
where $\mathbb{R}^n$ is considered with a specific inner product given by
$\langle z,z'\rangle_n^\beta=\sum_{k=0}^{n-1} e^{-\beta \lambda_k} z_k z'_k$
for all $z$ and $z'$ in $\mathbb{R}^n$. 

Once again, as stated by Corollary~\ref{eigen_communaute}, the vertices that
belong to the same perfect community have very close images by
$F_K^\beta$. However the embedding provided by $F_L$ (see equation
\eqref{F_L}) uses only a part of the spectrum and will therefore loose some
neighborhood informations. For example, vertices that belong to two different perfect communities can be indistinguishable. On the contrary, $F_K^\beta$ uses the whole eigen-decomposition but with a modified metric that contains non local information.

This approach is very flexible because the parameter $\beta$ permits to
control the degree of smoothing: a small value of $\beta$ regularizes heavily
and totally forbid to cluster together two vertices that are not directly
linked to each others whereas a large $\beta$ allows to cluster 
vertices that are not directly connected but share a large number of common
neighbors. This makes this kernel an attractive tool which is quite popular in the
computational biology area where it has been used with success to extract
pathway activity from gene expression data through a graph of genes
\cite{vert_kanehisa_B2003,scholkopf_tsuda_vert_KMCB2004}.

\section{Kernel SOM for clustering the vertices of a graph\label{som}}

\subsection{Motivations for the use of the SOM algorithm}
Our purpose is to provide a description of the graph by clustering its
vertices into relevant communities. However, clustering alone doesn't always provide 
a clear picture of the \emph{global} structure of a graph. As already
mentioned, on the one hand, perfect communities are easy to understand but
generally don't cover the whole graph, while, on the other hand, clusters that
are not perfect communities have complex relations one to another: a link
between two clusters hides a potentially complex link structure between
individuals in those clusters. A solution to circumvent those problems is
to cluster the vertices of the graph in a way that both leads to imperfect
communities but also takes into account relations between clusters.

To achieve these goals, one can leverage the topology preservation properties
of the Self Organizing Map (SOM). This algorithm, first introduced by Kohonen
\cite{kohonen_SOM2001}, is an unsupervised method that performs at the same
time a clustering and a non linear projection of a dataset. The SOM is based
on a set of \emph{models} (also called \emph{neurons} or \emph{units}) arranged
according to a low dimensional structure (generally a regular grid in one or
two dimensions). The original data are partitioned into as many homogeneous
clusters as there are models, in such a way that close clusters (according to
the prior structure) contain close data points in the original space.

The analysis of the vertices of a graph with a SOM will therefore provide a
type of relaxed communities (the clusters) arranged in a way that is consistent with
the link structure of the members of those communities, as long as the graph
structure can be turned into a topology that the SOM will preserve. 

\subsection{SOM for non vector data}
The standard SOM algorithm uses the euclidean structure of the data space and
therefore cannot be applied directly to vertices of a graph. As non vector
data arise naturally in many real world problems, adapted variants of the SOM
have received a lot of attention in the past ten years. It should first be
noted that the general structured data framework proposed in
\cite{hammer_etal_N2004} cannot be applied to vertices of a graph: the
framework is adapted to the case where each observation is a whole graph, not
to the one that focuses on the nodes of a single graph. 

A possible solution in this situation (explored in
\cite{villa_boulet_ESANN2007}) would be to use one of the variants of the
\emph{Median SOM} (also called the \emph{dissimilarity SOM}, see
\cite{ambroise_govaert_IFCS1996,kohonen_TR1996,kohonen_somervuo_N1998,kohonen_somervuo_NN2002,elgolli_rossi_conanguez_lechevallier_RSA2006}).
Members of this class of algorithms can be applied to any dataset on which a
dissimilarity measure can be defined: numerous dissimilarity measures for
graph nodes have been proposed for graph clustering (see
\cite{schaeffer_CSR2007}) and could therefore be used with a dissimilarity
SOM. Those SOM algorithms are based on a generalization of the notion of
center of mass called a generalized median (fast implementations are available
\cite{conanguez_rossi_elgolli_NN2006}). Another variant of the SOM for
dissimilarity data, based on mean field annealing, could also be used
\cite{graepel_etal_N1998,graepel_obermayer_NC1999}, as well as the recently
introduced relational topographic mappings
\cite{hammer_hasenfuss_TR2007,hammer_etal_WSOM2007}.

The solution proposed in this paper is to rely on a kernelized version of the
SOM: this is a natural choice in the sense that graphs are well described by
their Laplacian and the corresponding heat kernels. As shown in
\cite{villa_rossi_WSOM2007}, if the dissimilarity between objects is defined
via a kernel, the median SOM is a type of constrained kernel SOM. Moreover,
the constraints of the median SOM generally induce maps of lesser quality than
those obtained by the kernel version. Further links between both approaches
are outlined in \cite{hammer_hasenfuss_TR2007,hammer_etal_WSOM2007}.

In the proposed kernel approach, detailed in the next paragraph, the vertices are first
implicitly mapped into a feature space whose geometry reflects the graph
structure. This implicit mapping is performed via the so called ``kernel trick'', by using the diffusion kernel. Then, a batch SOM is applied in
this space to perform a nonlinear projection of the vertices and, at the same time, a
clustering, that will both respect the topology of the feature space and
therefore of the graph.

An alternative (and in fact quite similar) solution would be to rely on an
embedding, i.e., on an explicit mapping of the nodes of the graph to
$\mathbb{R}^{p}$, exactly as this is done in the spectral clustering approach
(see
\cite{cristianini_etal_NIPS2001,filippone_etal_PR2008,vonluxburg_TR2007}). Rather
than applying a $k$-means algorithm to the vector representation of the
vertices obtained via the mapping $F_L$ (see equation~\eqref{F_L}), one can
simply use a standard SOM. For the application studied in this paper (the
social network presented in Section~\ref{sectionMedievalNetwork}), this
solution performed poorly. While the overall organization of the obtained map
was good, the clusters were much more unbalanced: a large cluster contained
two third of the vertices, while other clusters were quite small (one or two vertices). This is not
very surprising as the heat kernel helps to distinguish between vertices that could seem similar if they are represented by the information restricted to the smallest eigenvectors of the Laplacian (as
explained in Sections~\ref{cutoff} and \ref{heat_kernel}).

Nevertheless, it should be noted that the important aspect of the proposed
method is to rely on an adapted variant of the Self-Organizing Map to perform
at the same time graph clustering and graph visualization. In the particular
application studied in Section \ref{sectionMedievalNetwork}, the heat kernel
(and therefore the batch kernel SOM) gives interesting results. In other
applications, better results might be obtained with other kernels, with
dissimilarities (via a dissimilarity SOM) or with embedding (via a standard
SOM). The numerous SOM variants provide a general framework for graph mining:
the present article explores only one of its possible concrete
implementation.

\subsection{Batch kernel SOM}\label{BatchKernelSOM}
Several kernelized version of the SOM have been proposed
\cite{macdonald_fyfe_ICKIESAT2000,andras_IJNS2002,villa_rossi_WSOM2007}. The
present paper uses a batch version of the kernel SOM proposed in
\cite{villa_rossi_WSOM2007,hammer_etal_WSOM2007}. An advantage of the batch kernel SOM, with respect
to the stochastic versions proposed before, is that the former generally
converges much faster than the latter. 

As stated above, the kernel batch SOM first maps the
original data into a high-dimensional Hilbert space $\mathcal{H}$ via a
feature map $\phi$. Then, the standard batch SOM is applied to the mapped
data. As with most kernelized algorithms, the mapping has not to be
explicitly carried out. The batch SOM can be rewritten in such a way to use
only the inner product of the Hilbert space: rather than defining
$\mathcal{H}$ and $\phi$, one has only to specify a \emph{kernel} $K$ on the
original data set, as this generates an associated Reproducing Kernel Hilbert
Space. 

Let us first describe the batch SOM on the mapped data. The prior structure
consists in $M$ neurons. The distance between neurons $i$ and $j$ in the prior
structure is denoted $h(i,j)$. It is transformed into a neighborhood function
via a decreasing function $R$, from $\mathbb{R}^+$ to $\mathbb{R}^+$, with
$R(0)=1$ and $\lim_{s\rightarrow +\infty} R(s)=0$. The influence of the grid
is annealed through time: at iteration $l$, the algorithm uses a function
$R^l$, based on $R$, that is more and more concentrated in $0$.

At iteration $l$, neuron $j$ is associated to a prototype (also called a code
book vector) $p^l_j$, chosen in $\mathcal{H}$, but constrained to be a linear
combination of the mapped data (as suggested in
\cite{macdonald_fyfe_ICKIESAT2000}), i.e.
\[
p^l_j=\sum_{i=1}^n \gamma^l_{ij} \phi(x_i).
\]
The batch kernel SOM is then given by Algorithm \ref{algoBatchKSOM}. 
\begin{algorithm}
\caption{The Batch Kernel SOM in feature space}\label{algoBatchKSOM}
  \begin{algorithmic}[1]
\STATE choose initial values for $\gamma_{ji}^0$ in $\mathbb{R}$
\STATE $p_j^0\leftarrow \sum_{i=1}^n \gamma_{ji}^0 \phi(x_i)$
\FOR{$l=1$ to $L$}
  \FOR[representation step]{$i=1$ to $n$}
  \STATE assign the observation $x_i$ to its closest neuron: 
\[
f^{l}(x_i)=\arg\min_{j=1,\ldots,M} \|\phi(x_i)-p_j^{l-1}\|
\]
  \ENDFOR
  \FOR[assignment step]{$j=1$ to $M$}
  \STATE update prototype $p_j$ according to
\[
p_{j}^l=\arg\min_{p=\sum_{i=1}^n \gamma_{i} \phi(x_i),\ \gamma\in\mathbb{R}^n} \sum_{i=1}^n R^l(h(f^l(x_i),j)) \|\phi(x_i) - p\|^2
\]
  \ENDFOR
\ENDFOR
  \end{algorithmic}
\end{algorithm}
It can be simplified by using the so called ``kernel trick'', which simply
consists in expressing operations in $\mathcal{H}$ solely via $K$. In
Algorithm \ref{algoBatchKSOM}, the value of
$\|\phi(x_i)-\sum_{j=1}^n\gamma_j\phi(x_j)\|$ has to be computed for any
linear combination $\sum_j \gamma_j \phi(x_j)$. This can be done via the following formulation
\[
\left\|\phi(x_i)-\sum_{j=1}^n\gamma_j\phi(x_j)\right\|^2=\|\phi(x_i)\|^2+\left\|\sum_{j=1}^n\gamma_j\phi(x_j)\right\|^2-2\sum_{j=1}^n\gamma_j\langle
\phi(x_i),\phi(x_j)\rangle.
\]
By definition, $\|\phi(x_i)\|^2=K(x_i,x_i)$ and $\langle
\phi(x_i),\phi(x_j)\rangle=K(x_i,x_j)$. Moreover
\[
\left\|\sum_{j=1}^n\gamma_j\phi(x_j)\right\|^2=\sum_{j=1}^n\sum_{j'=1}^n\gamma_j\gamma_{j'}\langle
\phi(x_{j}),\phi(x_{j'})\rangle=\sum_{j=1}^n\sum_{j'=1}^n\gamma_j\gamma_{j'}K(x_j,x_{j'}).
\]
Therefore, the assignment step of Algorithm \ref{algoBatchKSOM} simply reduces
to 
\[
f^l(x_i)=\arg\min_{j=1,\ldots,M} \sum_{u,v=1}^n \gamma_{ju}^{l-1}\gamma_{jv}^{l-1} K(x_u,x_{v}) - 2\sum_{u=1}^n \gamma_{ju}^{l-1} K(x_u,x_i),
\]
as $K(x_i,x_i)$ is fixed. Moreover, the solution of the minimization problem
of the representation step is given by $p_{j}^l=\frac{\sum_{i=1}^n
  R^l(h(f^l(x_i),j))\phi(x_i)}{\sum_{i=1}^n R^l(h(f^l(x_i),j))}$\footnote{this shows that choosing
  prototypes in the subspace spanned by the mapped data introduces in fact no
  constraint on them.} and therefore
the representation step can be simplified into
\[
\gamma_{ji}^l=\frac{R^l(h(f^l(x_i),j))}{\sum_{u=1}^n R^l(h(f^l(x_{u}),j)}.
\]
In practice, the $p^l_j$ don't have to be explicitly calculated, as $f^l$ is
computed directly from the $\gamma_{ji}^l$. It appears also clearly that
$\phi$ has not to be used and therefore that Algorithm \ref{algoBatchKSOM} can
be rewritten into the simpler Algorithm \ref{algoBatchKSOMKernelOnly}. 
\begin{algorithm}
\caption{The Batch Kernel SOM (simplified version)}\label{algoBatchKSOMKernelOnly}
  \begin{algorithmic}[1]
\STATE choose initial values for $\gamma_{ji}^0$ in $\mathbb{R}$
\FOR{$l=1$ to $L$}
  \FOR[assignment step]{$i=1$ to $n$}
  \STATE assign the observation $x_i$ to its closest neuron: 
\[
f^{l}(x_i)=\arg\min_{j=1,\ldots,M} \sum_{u,v=1}^n \gamma_{ju}^{l-1}\gamma_{jv}^{l-1} K(x_u,x_{v}) - 2\sum_{u=1}^n \gamma_{ju}^{l-1} K(x_u,x_i)
\]
  \ENDFOR
  \FOR[representation step]{$j=1$ to $M$}
  \STATE update prototype coordinates $\gamma_{ji}$ according to
\[
\gamma_{ji}^l=\frac{R^l(h(f^l(x_i),j))}{\sum_{u=1}^n R^l(h(f^l(x_{u}),j)}.
\]
  \ENDFOR
\ENDFOR
  \end{algorithmic}
\end{algorithm}

\subsection{Implementation details}
It is well known that the results of batch SOM strongly depend on the
initialization point, and also, but with a more limited scale, on the specific
implementation choice.

In the present paper, the prior structure is a bi-dimensional regular square
grid for which $h$ is given by the euclidean distance between the neurons. The
neighborhood function $R$ is a Gaussian function and generates a family
$R^l(x)=\exp(-x^2/T^l)$. The parameter $T^l$ is a temperature like parameter
which decreases over time in a geometrical annealing process (i.e.,
of the form $T^l=T^0\lambda^l$). The temperature is kept constant until
stabilization of the assignment step (i.e., until $f^l=f^{l-1}$) and then
decreased. This process is repeated until the temperature is low enough to
have $R^l(x)=1$ for $x=0$ and $R^l(x)\simeq 0$ for $x>0$: this ensures that the
algorithm will end with a final local organization behavior. This procedure is
quite standard for batch variants of the SOM. 

Two classical initialization strategies have been tested. In the first one,
the initial prototypes are randomly chosen among the original mapped data. In
practice, this is done by setting $\gamma_{ji}^0$ to $\delta_{i,k_j}$, where
$\delta_{u,v}=1$ if and only if $u=v$ and where $k_j$ is randomly chosen in
$1,\ldots,n$. The second initialization is a kernelization of the classical
Principal Component based method \cite{kohonen_SOM2001}. A Principal Component
Analysis (PCA) is conducted on the mapped data (this is therefore a kernel-PCA
\cite{scholkopf_etal_NC1998}) to discover the two principal directions. Then a
regular square grid is built on the two dimensional subspace spanned by those
directions. Coordinates of the vertices of the grid are used as initial values
for the prototypes: this can be done easily as the principal directions are
given as linear combinations of the mapped data.

\subsection{Comparing Maps}
A final problem is to choose the free parameters of the SOM, most importantly
the size of the grid and, in our case, the parameter $\beta$ of the diffusion
kernel. This latter parameter induces specific difficulties as the RKHS
associated to different values of $\beta$ use different metrics and cannot
therefore be directly compared. It is also well know that the final
quantization error
\[
\mathcal{E}=\sum_{i=1}^n\|\phi(x_i)-p^L_{f^L(x_i)}\|^2,  
\]
decreases with the number of clusters and fails also to measure topology
preservation. 

The problem of assessing the quality of a SOM has generated a large
literature. Among all the proposed topology preservation measures, the one
proposed by Kaski and Lagus in \cite{kaski_lagus_ICANN1996} seemed to be
well adapted to the considered problem. For the kernel SOM, the criterion is
given by 
\[
\mathcal{KL}=\frac{1}{n}\sum_{i=1}^n \left[\left\|\phi(x_i)-p^L_{f^L(x_i)}\right\|+ \min_{(j_0,\ldots,j_q)\in \mathcal{C}_i} \sum_{k=0}^{q-1} \left\|p_{j_k}^L-p_{j_{k+1}}^L\right\|\right],
\]
where $\mathcal{C}_i$ is the set of all paths in the prior structure starting
from $j_0=f^L(x_i)$ (the best matching unit for $x_i$), ending with $j_q$ the
second best matching unit for $x_i$ and such that $j_k$ and $j_{k+1}$ are
direct neighbors in the prior structure. The first part of the criterion is
exactly the quantization error, whereas the second term corresponds to a type
of continuity measure. This term is small when close points in the mapped
space have contiguous best matching units in the map. In the graph context,
this translates to the following statement: vertices that are close to each others in the feature space given by equation~\eqref{F_K} should be
mapped to close units on the map. It should be noted that even if the first
term of this criterion decreases with the size of the map, this is balanced by
the second term as a small quantization error cannot be achieved with very
close prototypes. The criterion can therefore be used to compare different
sizes for the map, even if it's likely to favor large maps. 

In addition to Kaski and Lagus' measure, the $q$-modularity
\cite{newman_PHE2003} was also considered. This graph clustering performance
criterion is defined by
\[
Q_\text{modul}=\frac{\sum_{j=1}^M (e_j - a_j^2)}{1-\sum_{j=1}^M a_j^2},
\]
where $e_j$ is the fraction of edges in the graph that connect two vertices in
cluster $j$ and $a_j$ is the fraction of edges in the graph that connect to
one vertex in cluster $j$. A high $q$-modularity means that vertices are
well clustered into dense subgraphs having few edges between them. The measure
is only based on the clustering result and can therefore be used to compare
e.g., two different values of the $\beta$ parameter.

\section{Mining a medieval social network}\label{sectionMedievalNetwork}
\subsection{Motivations}
In the French medieval society, peasants constitute 90\% of the whole
population. Despite this majority position, historic studies are mainly
concerned by the dominant classes (nobility and clergy) because peasants
left very few written documents compared to the well-educated part of the
population. As a consequence, historic studies on these periods often describe
an anonymous peasant community related to a master, a seignory or a church.

\parpic[r]{\includegraphics[width=4 cm]{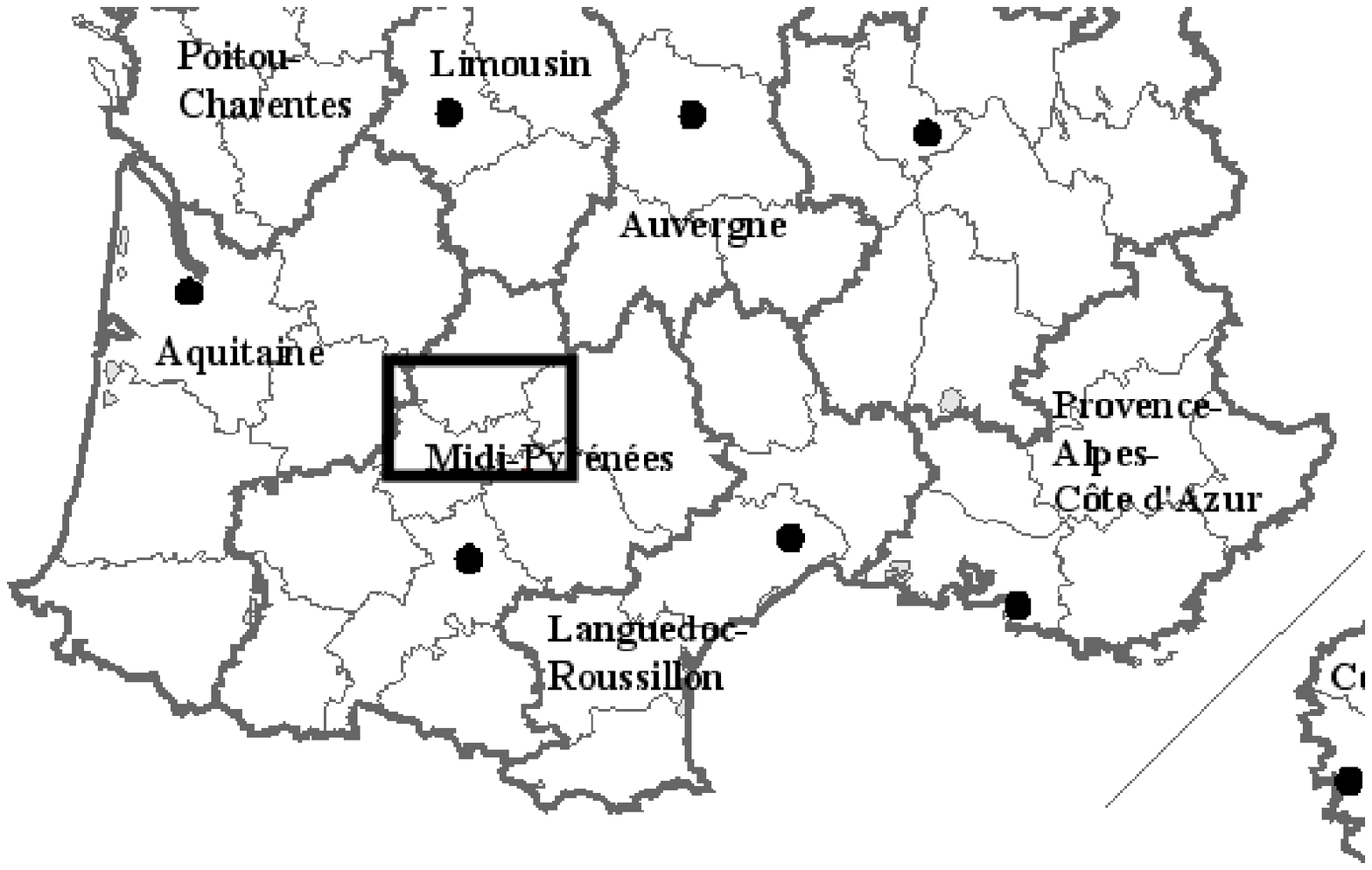}} In order to circumvent this difficulty, another approach has been pursued. The
main principle is to rely on agrarian contracts as a source of information
about social bounds between persons. We focus on a tiny geographical location
(several thousand hectares) for which a large documentation has been collected
(see \cite{hautefeuille_T1998} for a complete presentation). This
documentation is made from about 1000 agrarian contracts coming from about 10 villages located in the Castelnau-Montratier seignory which is a small area (about 30 km times 30 km) located in South West of France (Lot, in the rectangle on the right sided map). These contracts were first written between 1250 and 1350 and especially during the first 20 years of the XIV° century. After 1350, the documentation suddenly decreases because of the Hundred Years' War. All the contracts share common properties: they described land hiring, sales, legations and so on, they mention the name of the peasant (or the peasants) concerned by the transaction, the names of the lord and the notary with whom the peasants are related to, some of the neighbors of the peasants and various other informations (such as the type of transaction, the location, the date, and so on).

About 5000 additional similar contracts are still
to be recorded. The whole corpus, which is kept at Cahors (Archives Nationales
du Lot, France), has been totally rewritten during the XIX° century and is
therefore a very interesting source for historians as most of these types of
contracts have been destroyed, especially during the french revolution. A sociability
network of this peasant society can be constructed from the corpus. Because of
the size and the complexity of the obtained graph, automatic tools are needed
to understand it. The specific goal is to help historians to have a synthetic
view of the social organization of the peasant communities during the Middle
Ages.

\subsection{First description of the graph\label{desc}}
The corpus of agrarian contracts has partially been saved on a database. From
this database, a relational graph is built according to directives provided by
the historians and summarized below. Each vertex of the graph corresponds to
one person named in the contracts. First, nobles and notaries are removed
from the analyzed graph because they are named in almost every 
contracts: they are obvious central individuals in the social relationships and
could mask other important tendencies in the organization of the peasant
society. Then, two persons are linked together if: 
\begin{itemize}
	\item they appear in a same contract,
	\item they appear in two different contracts which differ from less than 15 years and on which they are related to the same lord or to the same notary.
\end{itemize}
The three main lords of the area (Calstelnau Ratier II, III and Aymeric de
Gourdon) are not taken into account for this last rule because almost all the peasants are
related to one of these lords. The links are weighted by the number of
contracts satisfying one of the specified conditions. Finally, 
the analysis is restricted to the largest connected component of the obtained
graph: it contains more than 80\% of its 
vertices.

\begin{figure}[ht]
	\begin{center}
		\includegraphics[angle=-90,width=7 cm]{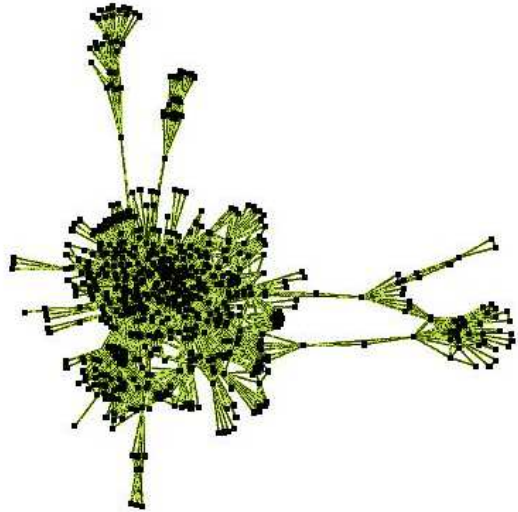}
		\caption{Representation of the medieval social network with force directed algorithm\label{graphe_init}}
	\end{center}
\end{figure}

This graph $\mathcal{G}$ has $615$ vertices and $4~193$ edges. The sum of the
weights is $40~329$, but almost 50\% of the edges have a weight $1$ and less
than 2\% have a weight greater than $100$. A simple representation of the
graph is given in Figure \ref{graphe_init} (this figure has been made by the
use of a force directed algorithm performed by the open source graph drawing
software Tulip\footnote{avaliable at
  \url{http://www.labri.fr/perso/auber/projects/tulip/}} \cite{auber_GDS2003} ).

\begin{figure}[htb]
	\begin{center}
		\includegraphics[width=7 cm]{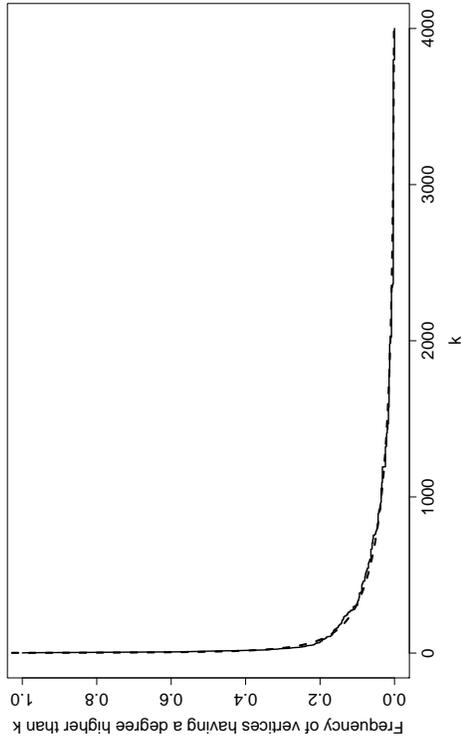}
		\caption{Cumulative degree distribution (solid) of the weighted graph fitting $P(k)= 1.12k^{-0.37}e^{-0.0009k}$ (dashed).\label{distr_degre}}
	\end{center}
\end{figure}

As this is frequently the case, the obtained social network is a small-world
network with low global connectivity and a high \emph{local connectivity}
\cite{watts_strogatz_N1998,watts_SWDNBOR1999}. Indeed the diameter of
$\mathcal{G}$ is $10$
and the mean of the shortest paths between two vertices is $3.9$. The local
connectivity, measured by averaging the density of subgraphs induced by the
direct neighbors of a vertex \cite{watts_SWDNBOR1999}, is $77\%$ whereas the
density of the graph is only $2.2\%$. The degree distribution also
obeys to standards (see
\cite{faloutsos_etal_ACMSIGCOMMCCR1999,newman_etal_PNASUSA2002,mossa_etal_PRL2002}):
the cumulative degree distribution for the weighted graph fits a power-law
with a fast decaying tail as shown in Figure \ref{distr_degre}. It follows
that the number of vertices having a degree $k$ is decreasing very fast (exponentially) when
$k$ increases and is not centered on a mean value: most of the peasants have a
small number of relationships and a tiny number of them have numerous
relationships.

\subsection{Clustering the medieval graph into perfect communities and rich-club}\label{subsec:medieval:perfect}

By the use of Theorem~\ref{vandenheuvel}, all the perfect communities of a graph $\mathcal{G}$ can be computed. They emphasize the main dense parts of the social networks but also discriminate individuals by their relationships (direct neighbors). 76 perfect communities were found in $\mathcal{G}$, most of them being very small (only 2 or 3 persons).

\begin{figure}[ht]
  \begin{center}
    \begin{tabular}{ccc}
      \includegraphics[width=5 cm]{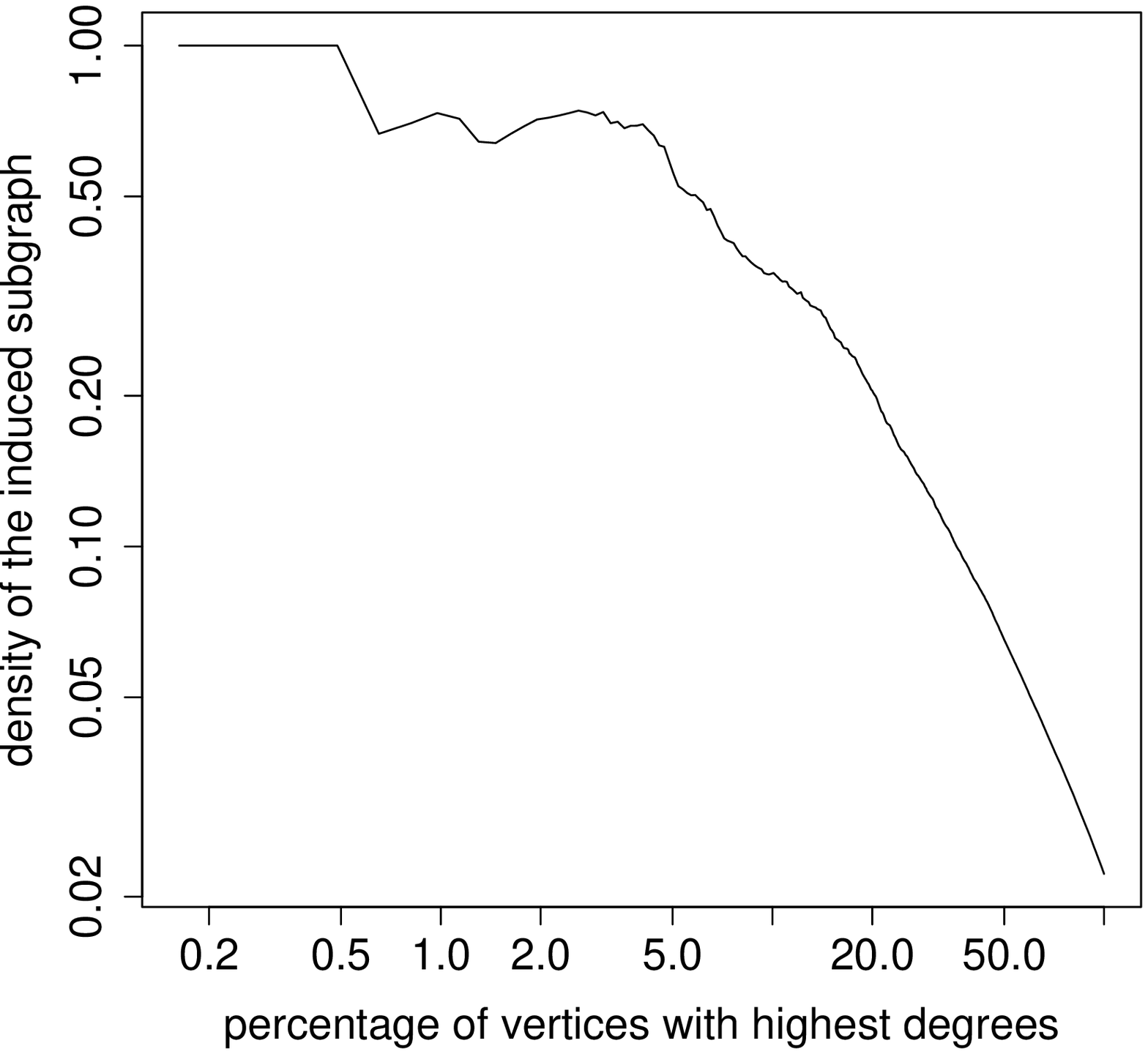} &\hspace*{0.5 cm} & \includegraphics[width=5 cm]{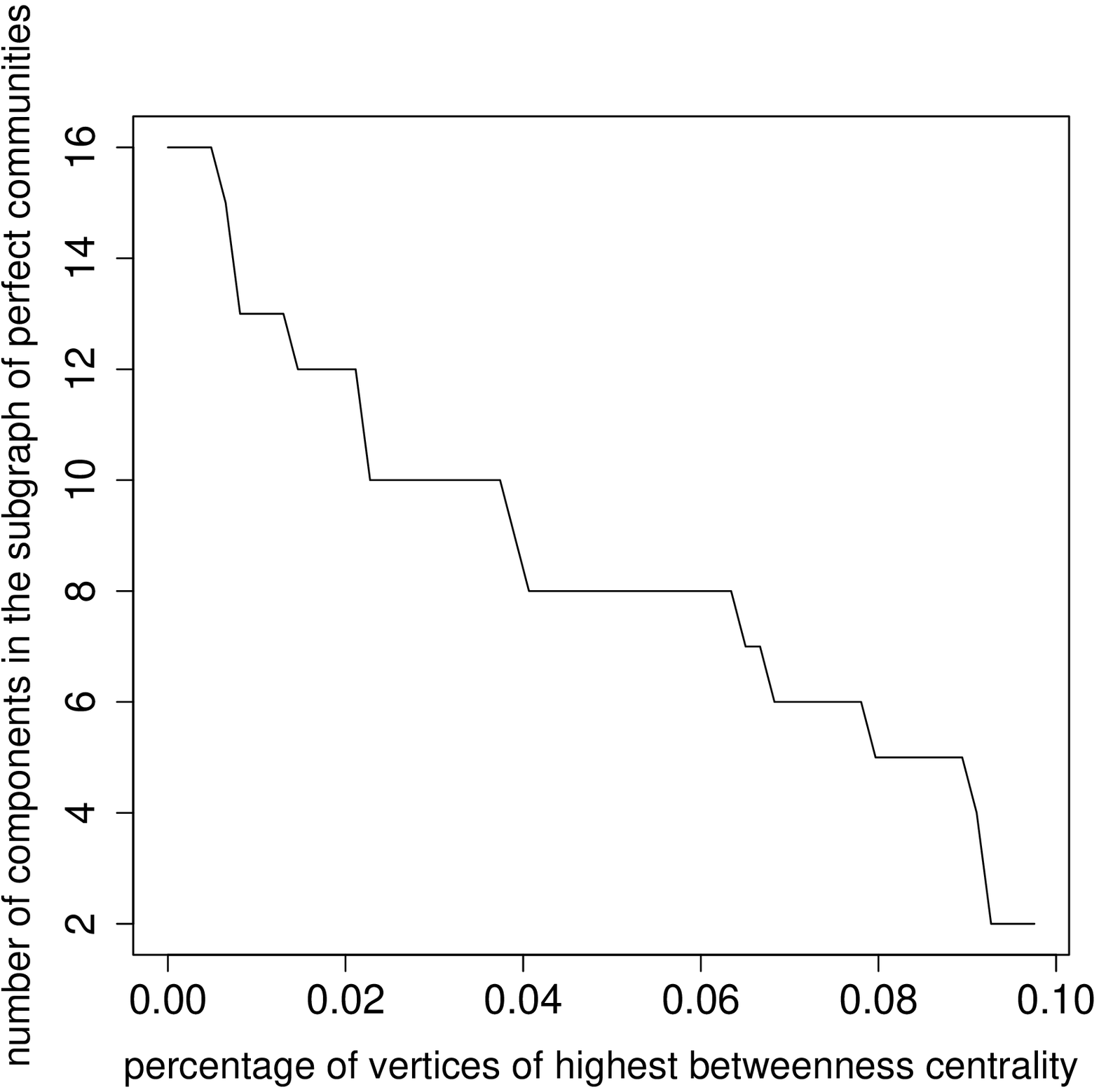}
    \end{tabular}
    \caption{Left: Density of the induced subgraph as a function
      of the number of highest degrees vertices (log scale) and
      Right: Number of components of the subgraph of perfect
      community and rich club as a function of the number of
      vertices with high betweenness measure
      added (in both cases, the number of vertices under consideration is given as a percentage of the total number of vertices in the graph).\label{riche_et_central}}
  \end{center}
\end{figure}

Then, as described in Section~\ref{algebric}, the rich-club and central vertices are extracted. The vertices in the rich club corresponds to the largest subgraph with highest degrees vertices having a diameter equal to 2. The rich club contains 3\% of the vertices of the whole graph which corresponds to 19 vertices. This subgraph has a high density as shown in Figure~\ref{riche_et_central}.
Central vertices are chosen to be the 4\% of the highest betweenness measure vertices of the whole
graph (i.e., 24 vertices). As show in Figure~\ref{riche_et_central}, it is a good compromise for this application: the
derived subgraph contains 8 components and a large number of vertices is
needed to decrease this number again. Moreover, except for one of them, these
components are tiny single perfect communities that won't be considered in the
following.

\begin{figure}[ht]
 	\begin{center}
 		\includegraphics[width=14 cm]{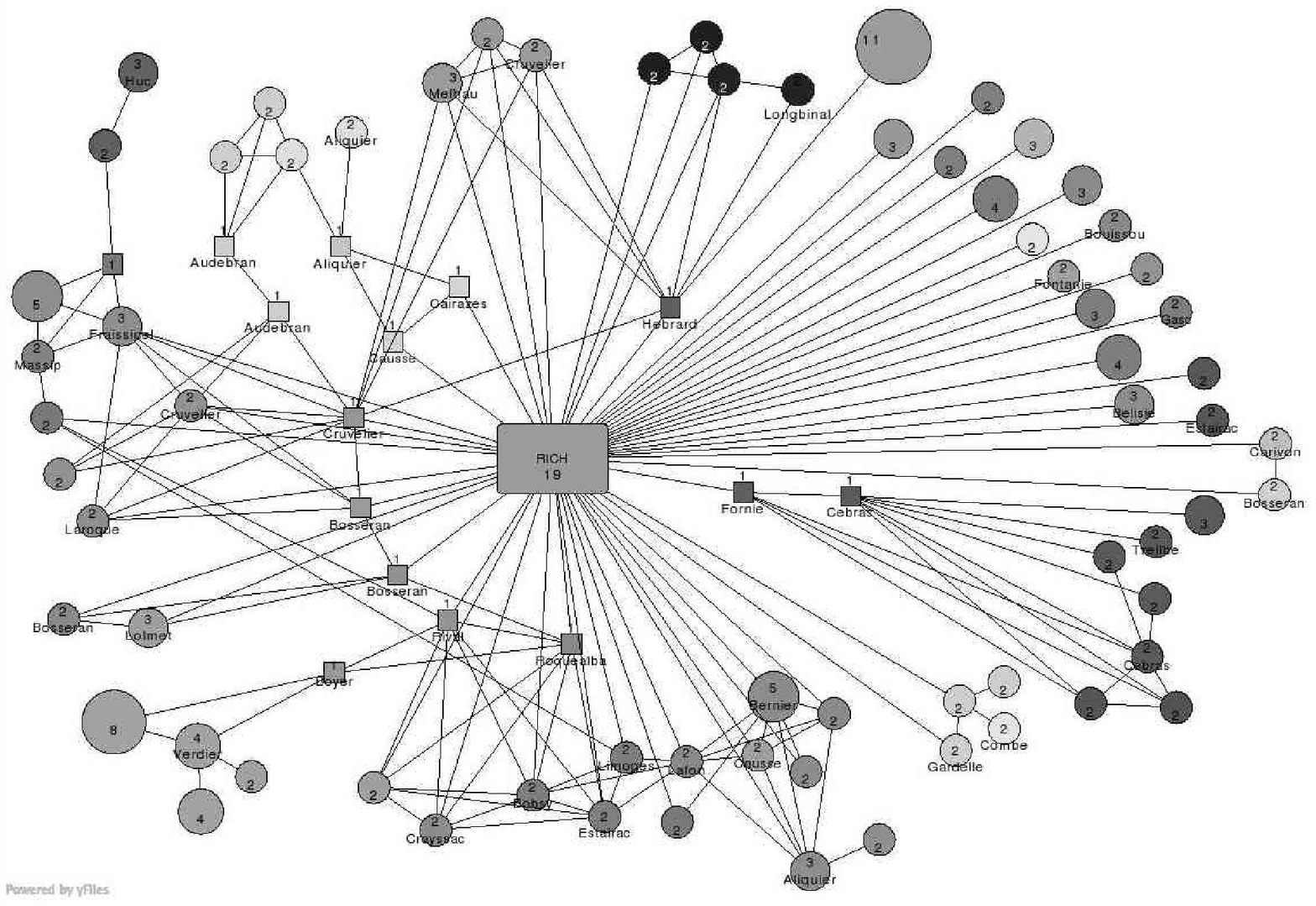}
 		\caption{Graph of the perfect communities (circles), the rich-club (rectangle) and central vertices (squares). Other details about the figure are given in the text.\label{fig_comm}}
 	\end{center}
 \end{figure}

Figure \ref{fig_comm}\footnote{This figure, and similar ones, have been made
  with the help of the free graph drawing software yED, available at
  \url{http://www.yworks.com/en/products_yed_about.htm}} provides a
representation of the perfect communities structure of the medieval social
network together with the rich-club and central vertices. The visual
representation of each perfect community has several features. The surface of
each disk is proportional to the size of the perfect community (i.e., to the
number of peasants in the perfect community) which is also recalled explicitly
by a number written inside the circle. The gray level of the disk encode the
mean date of the contracts in which the members of the community are involved
(from black, 1260, to white, 1340). In addition a family name is added when
the corresponding perfect community comes from a single family. The
communities are set at random positions but efforts have been done to
represent perfect communities that are linked by an edge at nearby
positions. Two perfect communities that are linked by an edge form a complete
subgraph but the peasants in this subgraph do not necessarily have the same
outside relationships; on the contrary, two peasants contained in the same
perfect community have exactly the same outside links. Links starting from
a community are therefore valid for all the members of this community. Seven
communities, that are still not connected with another perfect community, with
the rich club or with one of the vertices with a high betweenness measure,
were not considered for this representation.

\begin{figure}[ht]
	\begin{center}
		\includegraphics[width=9 cm]{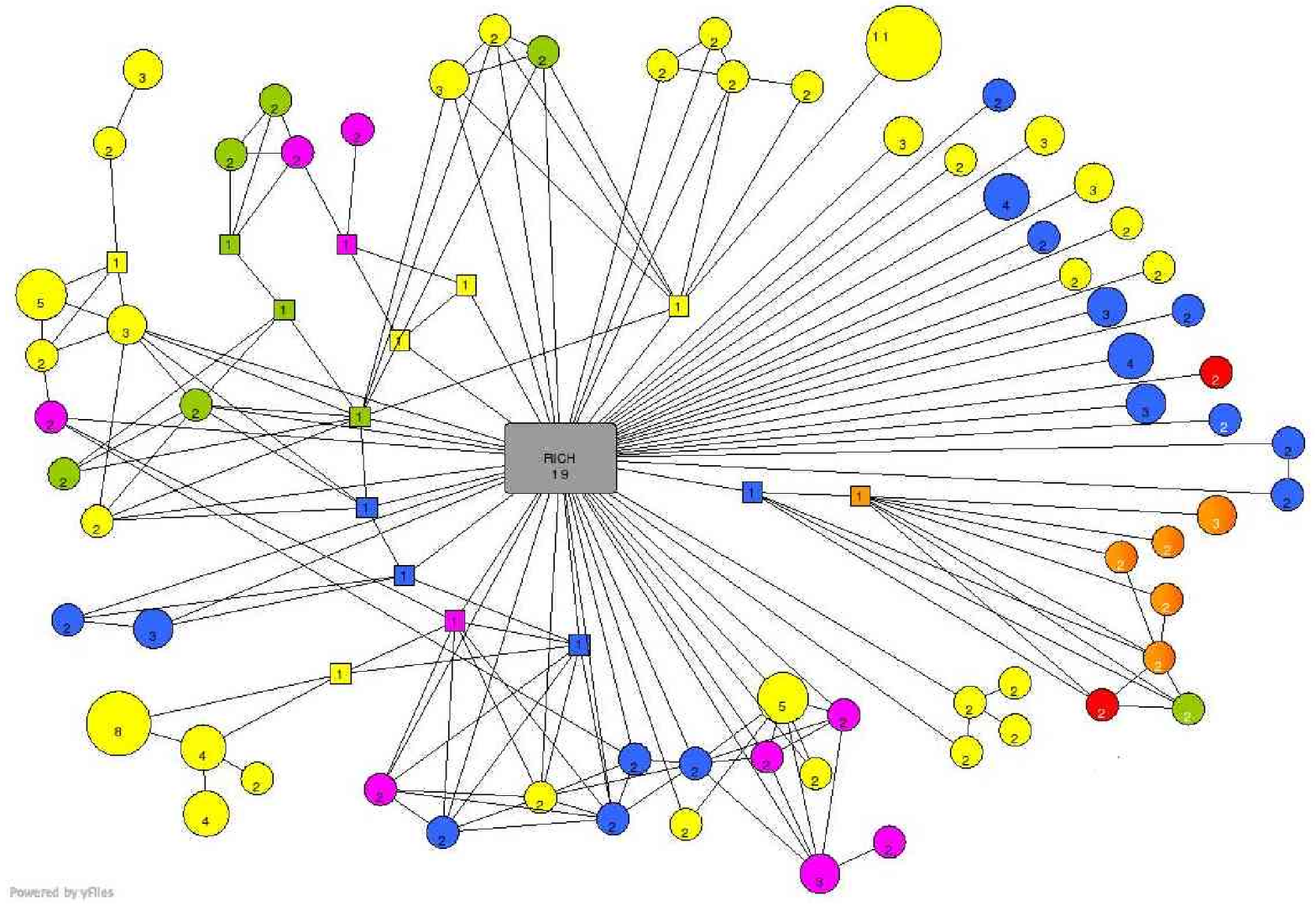}
		\caption{Graph of the perfect communities by geographical locations (yellow: Flaugnac, blue: Saint-Julien, green: Pern, pink: Cornus, red: Ganic and orange: Divilhac).\label{fig_comm_lieux}}
	\end{center}
\end{figure}

Figure \ref{fig_comm_lieux}\footnote{All colored figures are available on the
  publisher site.} provides an alternative representation of the
communities of the graph. Noting that the peasants of a perfect community
lived at the same geographical location, each perfect community is colored to
represent this location. The communities are set at the same positions as in
Figure~\ref{fig_comm}.

Of course, these representations have to be interpreted with care; some of
their properties could induce interpretation biases. One of the main
limitation is that only a part of the vertices of the original graph is
represented on it (35 \% of them). Moreover, the respective positions of the
perfect communities can be relevant (if their are linked) or not (if they are
not). Nevertheless, these ``maps'' of the graph allows to understand some
important facts about the medieval society. First of all, the ``small world''
structure of the graph is emphasized by the star shaped structure of the
perfect communities around the rich-club: some persons seem to belong to
small groups (a perfect community or linked perfect communities) which are only
related to each others by the way of the main individuals (the rich-club or
peasants with a high betweenness measure).

Then, family links seem to have a great importance in the medieval society as
all individuals in the perfect communities often share the same family name
(this is the case for 30 perfect communities) but geographical proximities are
even more important: as shown in Figure~\ref{fig_comm_lieux}, all the perfect
communities have homogeneous locations and very often, linked perfect
communities also share the same geographical location. Finally, it appears
that persons with a high degree of betweenness share the same geographical
location as the perfect communities they are linked to. These individuals can
be seen as peasants making the link between several villages or between a
village and one of the central person from the rich-club.

\subsection{Mapping the medieval graph with the SOM}
The social network was analyzed with the batch kernel SOM as follows. The
parameter $\beta$ varied between $0.01$ to $0.05$. Values above $0.05$ lead to
instability in the calculation of the diffusion matrix in the sense that the
obtained kernel is no more positive. Values smaller than $0.01$ lead to hard
clustering (the diffusion matrix is close to the identity matrix) that are not
relevant (see \cite{villa_rossi_WSOM2007}).

For a fixed kernel (i.e., a value of $\beta$), all squared maps from
$5\times5$ to $10\times10$ were tested as prior structures of the grid. For
each prior structure, several random initial configurations, the kernel PCA
based initial configuration and several initial temperatures were compared via
Kaski and Lagus' quality measure, leading to the selection of a single final
map for each size (it should be noted that kernel PCA, associated to an
optimal choice of the initial temperature, leads to much better results than
random initialization).

In terms of $q$-modularity, the quality of clustering results increases with
the value of $\beta$ (for almost all map sizes). As a consequence, the value
of $\beta=0.05$ was selected. Among the 6 maps built with this kernel, those
of size $6\times 6$ and $7\times 7$ are the most interesting. The first one
has the highest $q$-modularity, whereas the second has the smallest value of
$\mathcal{KL}$ criterion together with a high value of the $q$-modularity. 

\begin{figure}[ht]
	\begin{center}
		\begin{tabular}{cc}
			\includegraphics[width=7 cm]{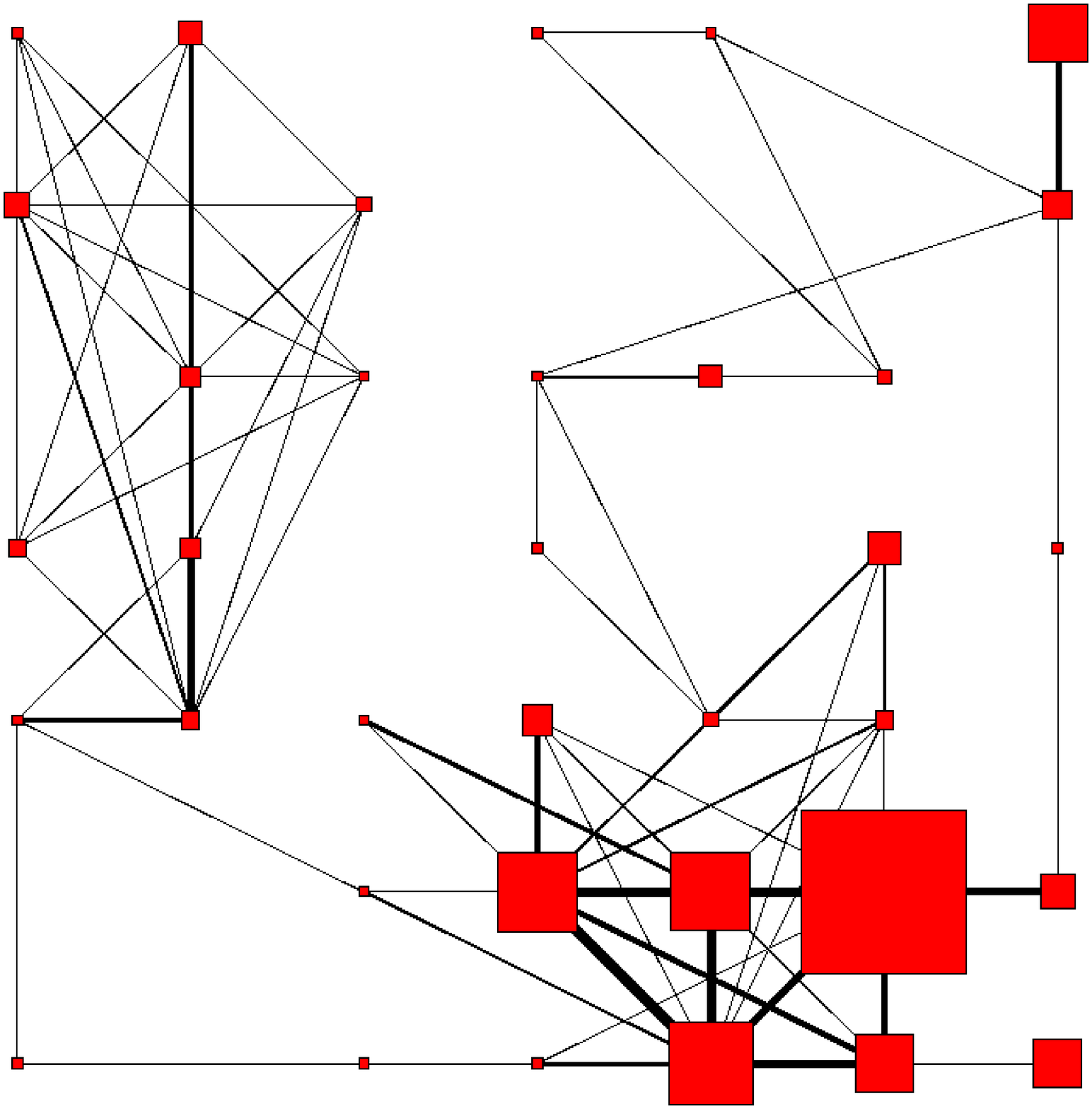} & \includegraphics[width=7 cm]{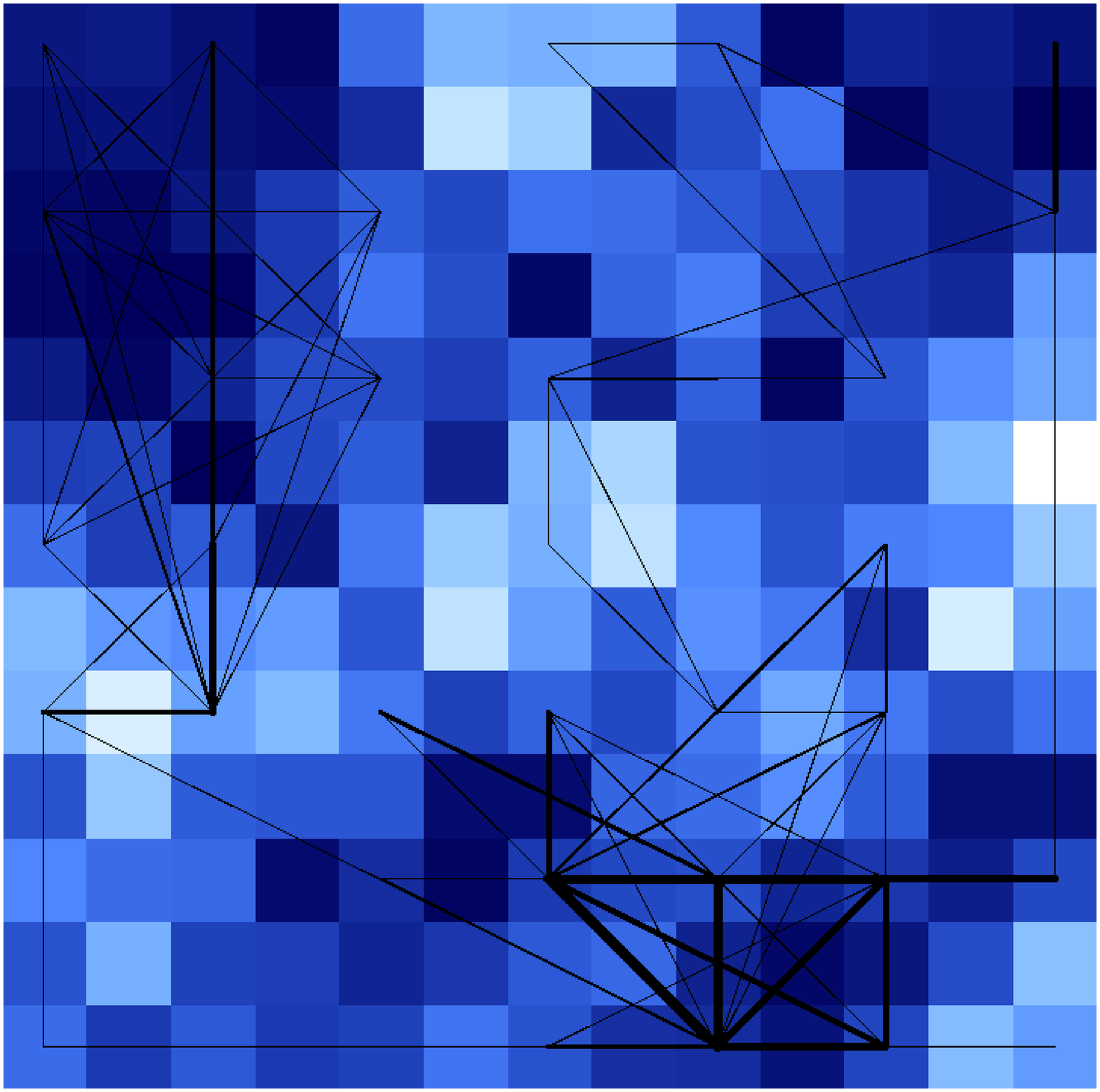}
		\end{tabular}
		\caption{Final self-organizing map ($7\times7$ square grid)\label{fig_som}}
	\end{center}
\end{figure}

We decided to focus on the $7\times 7$ map as it seems to be the most
interesting. It contains 35 non empty clusters and is given in the left part
of Figure~\ref{fig_som}. In this graphical representation, the surface
occupied by a square is proportional to the size of the corresponding cluster,
while the width of the connection between two squares is proportional to the
total weight of the edges connecting vertices of the two clusters. 

The right part of Figure \ref{fig_som} is the U-matrix
\cite{ultsch_siemon_INNC1990} of the map. It visualizes distances between
prototypes (in the mapped space): dark colors correspond to close prototypes
and light colors to a large distance between the corresponding prototypes.

The map is divided into three dense subparts: top-left, top-right and
bottom-right. The number of edges is small between these three parts and much
more dense inside the clusters of the same part, which seems to be
relevant. The most dense part of the map is the bottom-right one: one of its
clusters contains 255 vertices, which represents more than one third of the
whole graph. This part is connected to the two others which are not connected
to each others.

\begin{figure}[ht]
	\begin{center}
		\includegraphics[width=7 cm,angle=180]{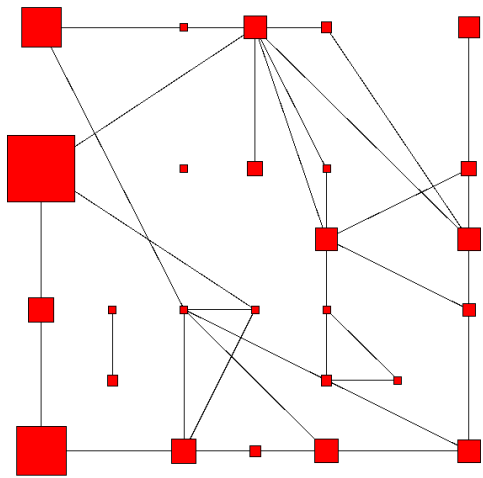}
		\caption{Self-organizing map of the main cluster\label{fig_som_gc}}
	\end{center}
\end{figure}

As the largest cluster still seems to be too large to be relevant, another
batch kernel SOM was constructed on the subgraph induced by the vertices of
this cluster; this methodology is known as a \emph{hierarchical feature maps}
\cite{miikkulainen_CS1990}. As before a $7\times7$ map is selected. It is
represented in 
Figure~\ref{fig_som_gc}. 
This map seems to be well connected except 3 little clusters that are not
connected to the rest of the map. The final map is sparse and well organized.
Once again, the main cluster of this map contains a high number of vertices, 81,
which represents almost one third of the whole subgraph.

Such a phenomen reflects
the cumulative degree distribution described in Section~\ref{desc}.
An analysis of the degree distribution on the map of Figure~\ref{fig_som} shows that the 10\% of the
vertices having the highest degrees are all
clustered in three clusters of the bottom right part of the map (but not in
the largest one, GC\footnote{The way the clusters are referenced is indicated in Figure~\ref{col_comm}.}). Then, looking at the
degrees of the subgraph made from the vertices in cluster GC, we see that, once again, their cumulative distribution is a power-law cumulative distribution
but with another scale (the density of this subgraph is 5 times smaller than the one of the
initial graph). The same phenomenon occurs in the subgraph GC: the 10\%
of the vertices that have the highest degrees are all assigned to the three
mainly connected clusters of Figure~\ref{fig_som_gc} and the largest cluster of this subgraph (GC20\footnotemark[6]) also has a density
4 times less than the whole subgraph GC.

\subsection{Historical properties of the self-organizing map}
\begin{figure}[ht]
	\begin{center}
		\includegraphics[width=9 cm]{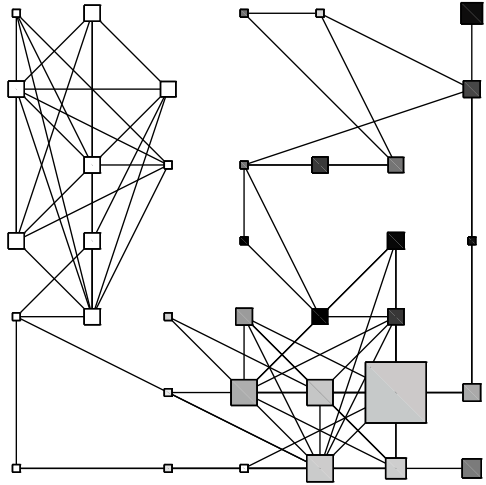}
		\caption{Mean date for each cluster from black, 1260, to white, 1340\label{dates}}
	\end{center}
\end{figure}

The analysis mimics what has been done for the perfect communities, starting
with the distribution of the dates on the map. The mean date for each cluster
is depicted on Figure~\ref{dates} using a gray scale. Each cluster has a small
standard deviation; clusters having the highest standard deviations are the
most connected clusters of the bottom right part of the map.
The three parts of the Kohonen map emphasized by the U-matrix have homogeneous
dates: top right is the oldest part, bottom right have middle dates and top
left, the most recent dates. The clusters are continuously connected to each
others by the date, in the sense that connexion clusters have intermediate
dates. The organization provided by the SOM is therefore relevant. But, since
the studied period is only 100 years long, this also seems to show that
various generations (sons, fathers, grand-fathers,\ldots) are not highly
mixed; particularly, the earlier part of the map is only connected to the rest
by a very few number of individuals (1 to 3).

The geographical locations of the persons belonging to the same cluster are
generally homogeneous. More precisely, they are exactly the same for peasants
belonging to the same little cluster and the largest clusters often have a
dominant geographical location but also contain peasants that don't live in
this geographical location. The family names are generally not the same for
peasants in the same cluster, with some exceptions as, for example, cluster 10\footnotemark[6]
which corresponds to ``Aliquier'' family, just as one of its closest cluster,
11\footnotemark[6]. Thus, as already mentioned in the analysis of the perfect communities,
geographical proximities seem to have a main role in the peasant's
relationships.

\subsection{Comparison with the work on perfect communities}
\begin{figure}[ht]
	\begin{center}
		\begin{tabular}{c}
			\includegraphics[width=14 cm]{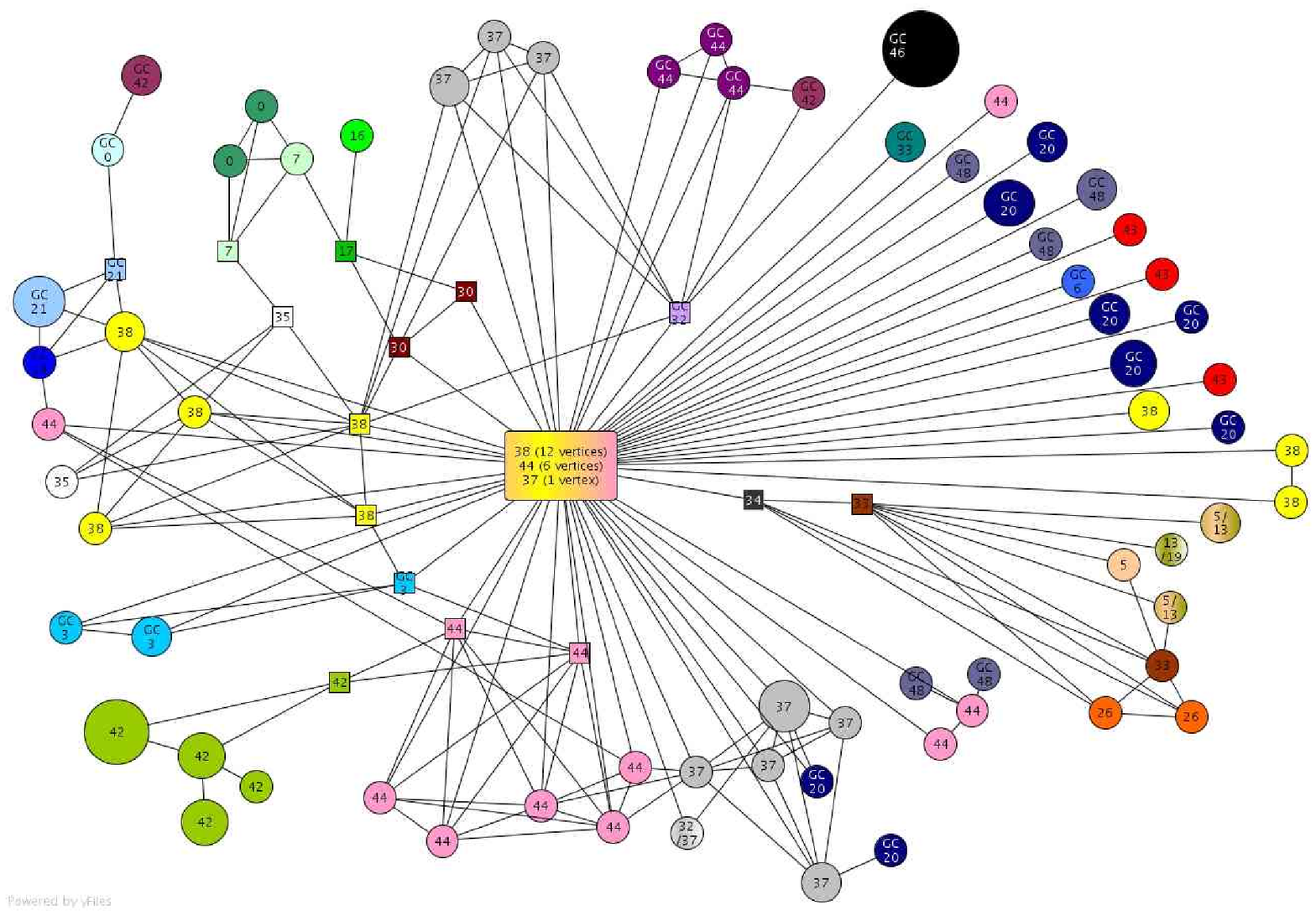}\\
			\begin{tabular}{cc}
				\includegraphics[width=8 cm]{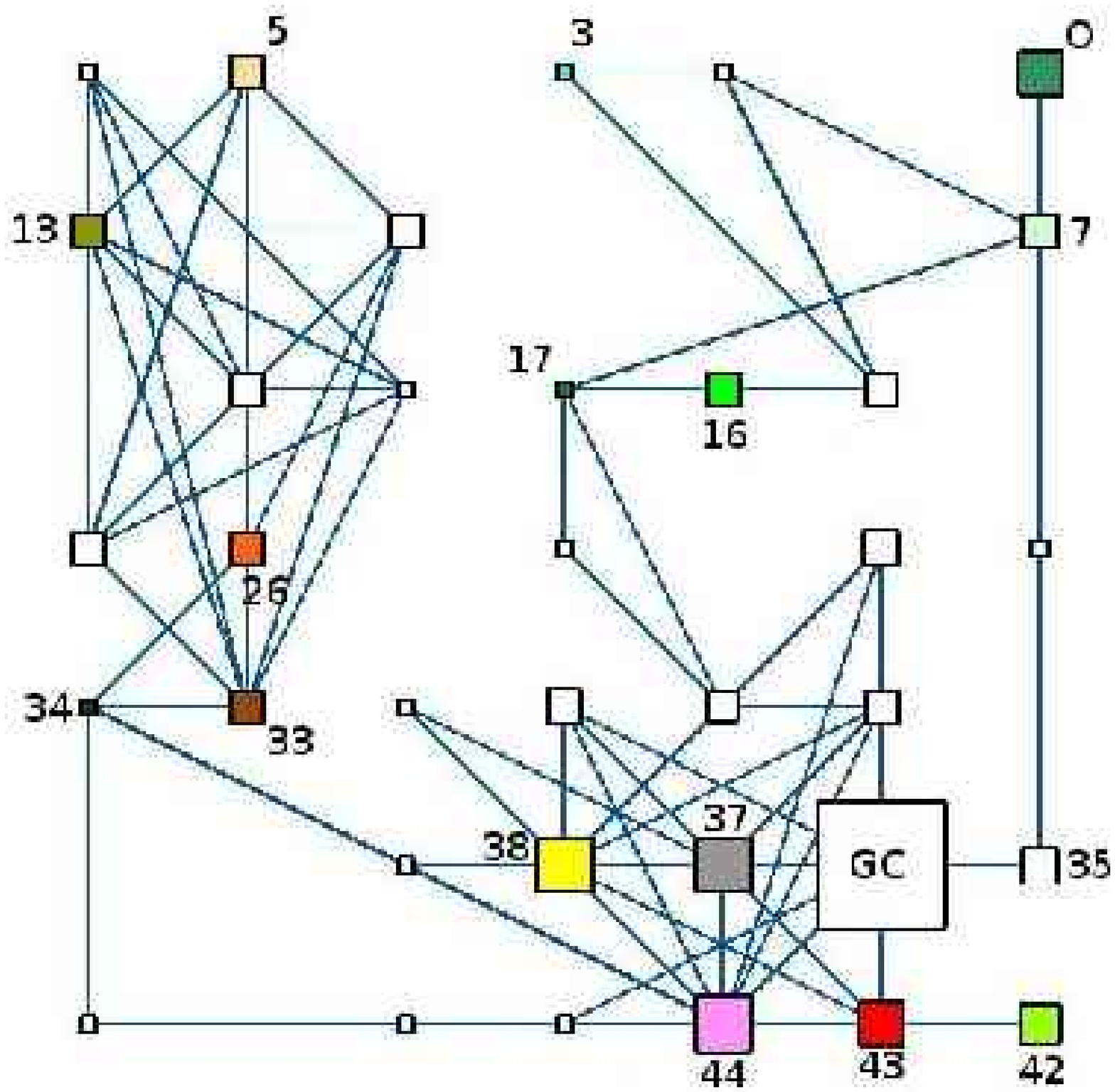} &
				\includegraphics[width=4 cm]{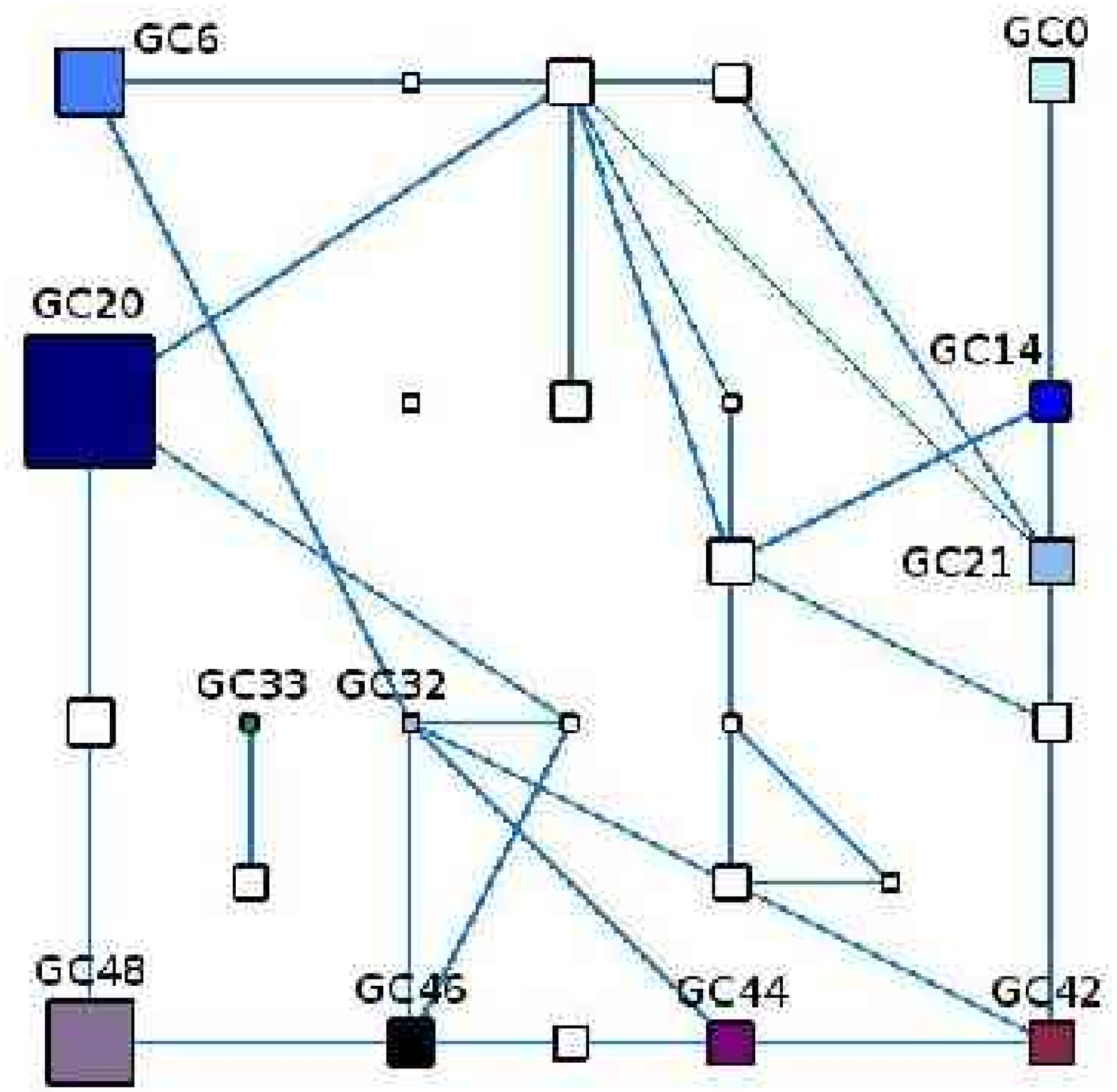}
			\end{tabular}
		\end{tabular}
		\caption{Comparison of the representation of the graph through perfect communities and self-organizing maps (left : of the whole graph; right : of the largest cluster of the initial SOM. More details about this figure are given in the text)\label{col_comm}}
	\end{center}
\end{figure}

A comparison of the self-organizing map with the perfect community
representation (Figure~\ref{fig_comm}) is provided by Figure~\ref{col_comm}:
the vertices that belong to the same perfect community are almost always in
the same cluster of the self-organizing maps (except for three small perfect
communities). To study the reverse mapping, an arbitrary color was assigned to
each cluster that contains at least one perfect community and then used to
color the same way the perfect communities of Figure~\ref{fig_comm}. The
number assigned to each perfect community is the number of the cluster in one
of the two maps (prefixed by ``GC'' for the clusters of the largest cluster of
the initial SOM).

Figure~\ref{col_comm} emphasizes the great similarity between the two
approaches: perfect communities that share links often belong to the same
cluster of the self organizing map. A similar remark can be made about
peasants with a high betweenness measure: they are often assigned to the
same cluster as the perfect community that they link to the rich-club.
Moreover, perfect communities that share a link or that are linked to the same
peasant with a high betweenness measure but have different colors often belong
to nearby clusters on the SOM: this is the case, for example, for clusters 44
and 37, for clusters 0 and 7, for clusters 26, 33 and 34, etc. It is also
interesting to note that some of the persons having a high betweenness measure
also have an important position on the SOM: for example, peasants 17 and 34
are emphasized by the fact that they link the bottom right part of the map
with the top right part and the top left part, respectively.  All these
similarities are evidence that there is a strong consensus between both
approaches and, as a consequence, that they offer a realistic representation
of the organization of the peasant society in the Middle Ages.

Nevertheless, there are also some interesting differences between the two
approaches. First of all, it is suprising that the rich-club is separated in
several clusters (37, 38 and 44) in which some perfect communities can also be
found. Arguably, these three clusters are very close on the SOM and have
strong connectivity (depicted by the tick lines between them). Moreover, the three clusters of the SOM
correspond to different geographical locations: cluster 37 contains a majority
of peasants living in ``Cornus'', cluster 38 and 44 in ``Saint Julien''. In
addition, Clusters 38 and 44 also contains peasants that have different family
names: ``Belisie'', ``Bernier'', ``Bosseran'', ``Cruvelier'', ``Laroque'',
``Ratier'' and ``Sirven'' are found several times in cluster 38 but none in
cluster 44 and ``Amilhau'', ``Camberan'', ``Labarthe'', ``Limoges'', ``Rival''
and ``Tessendie'' are found several times in cluster 44 but none in cluster
38. However, families ``Estairac'' and ``Fague'' are well represented in both
clusters 38 and 44. It is therefore not very clear whether the separation of
the rich-club into three clusters is relevant or not. An advantage of the
rich-club approach over the SOM based analysis is to emphasize the members of
this group who clearly have a special social role, while there is nothing very
specific about the corresponding clusters in the map. 

It appears also that some perfect communities share the same color whereas
they don't seem to be ``close''. Sometimes, this is due to the fact that the
positions of these perfect communities are partially random despite the fact
they are linked to each others (this is the case, for instance, for the pink
group of perfect communities 44 at the bottom of the figure and the perfect
community of the same color at the left part of it). Sometimes, this can be
explained by links that are not represented on the figure: for example,
cluster 38 is separated into three groups of perfect communities that are not
linked to each other but these groups share some common relationships with
vertices in cluster 38 in the rich-club. However this argument is less
convincing to explain why cluster 37 contains two groups of perfect
communities. Finally, in some cases, there is no simple reason to explain why
several perfect communities are grouped in the same cluster: for example, GC20
is still a large cluster that contains several perfect communities that are not
linked to each others on the perfect communities representation.

\subsection{Conclusion}\label{conclusion}
The remarks made about the similarities and differences between the two
approaches show that they can both provide elements to help the historians to
understand the organization of the medieval society. Moreover, they have
distinct advantages and weaknesses. 

On the one hand, representing the graph through its perfect communities
induces the question of the way these communities have to be represented in a
two-dimensional space, even if the restrictive definition chosen for
communities partly reduces this problem. This question is difficult (and
related to the field of graph drawing) but of a great importance to avoid
interpretation bias. For this point, the kernel SOM can appear as an
alternative that provides a notion of proximity, organization and even distance
between the communities. Moreover, kernel SOM allows to organize all the
vertices of the graph and not only the vertices that belong to a perfect
community.

On the other hand, the links inside and outside the clusters of the kernel SOM
are not clear: some of the vertices in a cluster can have no edge in common
with the other vertices of the cluster (it is the case, e.g., for one of the
cluster of the largest cluster as is shown by Figure~\ref{fig_som_gc}) and two
vertices in the same cluster are not necessarily related to the same vertices
outside the cluster. These two facts seem to show that kernel SOM probably
provides a better macroscopic view of the graph, whereas the perfect community
approach is more reliable for local interpretations: as the definition of a
perfect community is restrictive, it emphasizes very close social groups that
share the same geographical location and also often the same family name. The
interpretation of such social groups is then easier. 

It should be noted that in both cases, the social and historical analysis is
only facilitated by the algorithms rather than somehow being automated. In a
sense, the problem of understanding the social network is simply pushed a
little bit further away\footnote{The authors are grateful to one reviewer for
  pointing this out.} by the methods, especially in the case of the kernel
SOM. Figures \ref{fig_som} and \ref{fig_som_gc} for instance give broad
pictures of the social network, but a more detailed analysis is needed to
extract knowledge from the network. One of the interesting aspect of the
combined methodology proposed in the present paper is to help this detailed
analysis.

To go further, an open question is the way the parameters of the kernel SOM
have to be chosen and especially the size of the map, or, in the same spirit,
how deep a hierarchical analysis should be conducted on a large cluster. This
question is related to finding a relevant size for each community. The perfect
community approach can help driving this work by providing an idea of the
relevance of a given cluster, as we emphasized for cluster GC20.

Conversely, kernel SOM could also help to provide a more realistic
representation of the perfect communities in creating a drawing algorithm that
can also take into account the distances between clusters in the SOM. This
question is currently under development.

\section{Acknowledgments}
The authors would like to thank Florent Hautefeuille, historian (FRAMESPA,
Université Toulouse Le Mirail, France) for giving us the opportunity to work
with this interesting database and for spending time to explain us its
historical context. We also want to thank Fabien Picarougne and Bleuenn Le Goffic (LINA,
Polytech'Nantes, France)  for managing the database registration and Pascale Kuntz for her expertise in graph vizualisation. We finally
thank the anonymous reviewers for their detailed and constructive comments
that have significantly improved this paper.

\end{document}